\documentclass[aps,prl,reprint,floatfix,showpacs,longbibliography,superscriptaddress,nofootinbib]{revtex4-2}
\usepackage{graphicx}
\usepackage{amsmath}
\usepackage{amssymb}
\usepackage{amsfonts}
\usepackage{bm}        
\usepackage[mathscr]{eucal}
\usepackage[colorlinks, linkcolor=red,citecolor=blue,urlcolor=blue]{hyperref}
\usepackage[normalem]{ulem}
\usepackage[all]{hypcap}
\usepackage{multirow}
\usepackage{nicefrac}
\usepackage[dvipsnames,usenames,table]{xcolor}
\usepackage{soul,xcolor}
\usepackage[toc,page]{appendix}
\usepackage{dsfont}

\def\@fnsymbol#1{\ensuremath{\ifcase#1\or *\or \dagger\or \ddagger\or
   \mathsection\or \mathparagraph\or \|\or **\or \dagger\dagger
   \or \ddagger\ddagger \else\@ctrerr\fi}}

\begin{document}
\setstcolor{red}
\title{Robust Spin Polarization by Adiabatic Dynamical Decoupling}

\author{Soham Pal\textsuperscript{*}}
\affiliation{Cavendish Laboratory, University of Cambridge, JJ Thomson Avenue, Cambridge CB3 0HE, United Kingdom}

\author{Oliver T. Whaites\textsuperscript{*}}
\affiliation{Department of Physical Chemistry, University of the Basque Country UPV/EHU, Apartado 644, 48080 Bilbao, Spain}
\affiliation{EHU Quantum Center, University of the Basque Country UPV/EHU, Bilbao, Spain}
\affiliation{Department of Physical and Astronomy, University College London, Gower Street, London WC1E 6BT, United Kingdom}

\author{Wolfgang Knolle}
\affiliation{Leibniz Institute of Surface Engineering (IOM), Permoserstraße 15, 04318 Leipzig, Germany}

\author{Tania S. Monteiro}
\affiliation{Department of Physical and Astronomy, University College London, Gower Street, London WC1E 6BT, United Kingdom}

\author{Helena S. Knowles\textsuperscript{$\dagger$}}
\affiliation{Cavendish Laboratory, University of Cambridge, JJ Thomson Avenue, Cambridge CB3 0HE, United Kingdom}

\date{\today}

\begin{abstract}
High-fidelity multi-qubit initialization is vital for quantum simulation, quantum information processing (QIP), and quantum sensing. In diamond platforms, nuclear spin registers can be initialized through polarization transfer from a nearby electronic spin whose high gyromagnetic ratio enables efficient dynamical nuclear polarization (DNP). These hybrid systems are typically controlled using diabatic spin rotations, which require precise knowledge of all system parameters. Adiabatic DNP protocols on the other hand have less strict requirements and could enable robust and high fidelity spin transfer. However, due to the slow adiabatic sweeps and limited electron spin coherence times, this approach has remained inaccessible.  
Here, we demonstrate adiabatic pulsed nuclear spin polarization at room temperature in diamond. We achieve enhanced polarization efficiency, a broad resonance window, and improved tolerance to hyperfine coupling uncertainties relative to conventional diabatic pulsed protocols. We also show how this approach can benefit the initialization of spin clusters. These results set the scene for enhanced qubit initialization in solid state through adiabatic pulsed driving, with applications in solid-state quantum sensor and quantum memory technologies.
\end{abstract}

\maketitle 

\footnotetext{Authors equally contributed to work.}
\footnotetext{Corresponding Author: hsk35@cam.ac.uk}


\paragraph{Introduction:}
Solid-state defects with an electronic spin degree of freedom have emerged as versatile platforms for quantum computing \cite{bradley2019ten,west2019gate,abobeih2022fault,jelezko2004observation}, quantum networks \cite{gao2015coherent,awschalom2018quantum}, quantum simulation \cite{randall2021many,van2021quantum} and quantum sensing\cite{zhou2020quantum,aslam2017nanoscale}. Among these, nitrogen vacancy (NV) centers in diamond offer room-temperature optical initialization, spin-state readout, and long spin coherence times, combined with the availability of nearby coupled $^{13}\text{C}$ nuclear spins.
Coherent control and Dynamical Decoupling (DD) techniques, such as Carr-Purcell-Meiboom-Gill and XY-8, can be employed to reach second-scale spin coherence times \cite{bar2013solid,abobeih2018one}. They also allow for engineering interactions with coupled nuclear spins. When tuned to nuclear spin resonances, these DD sequences enable efficient indirect nuclear spin control and high-fidelity nuclear spin readout \cite{taminiau2012detection}, establishing hybrid NV–$^{13}\text{C}$ systems as a leading candidates for quantum information processing (QIP) applications.
A critical requirement for QIP is the initialization of qubits into well-defined states \cite{steane1998quantum}. While NV electronic spins can be optically initialized with high fidelity, the associated nuclear spins remain maximally mixed at room temperature. DNP transfers polarization from the electronic spins to the nuclear spins. In particular, pulsed DNP methods, such as PulsePol \cite{schwartz2018robust}, exploit hyperfine interactions to engineer flip-flop transitions for this purpose.
Alternatively, adiabatic state transfer, widely employed in atomic physics, achieves complete state transfer through a slow sweep of a control parameter such as microwave excitation (MW) detuning or external magnetic field \cite{vitanov2001laser}. However, the slow nature of traditional adiabatic sweeps often conflicts with limited NV coherence times ($T_2^*$) \cite{zhou2017accelerated}, making adiabatic protocols inaccessible. Strategies to mitigate decoherence by using Hamiltonian engineering to combine DD with adiabatic sweeps have been theoretically proposed \cite{quiroz2012high,witzel2015multiqubit,villazon2021shortcuts}. A new approach, AdPulse \cite{whaites2022adiabatic}, incorporates pulsed DD into an adiabatic sweep by using the Floquet phase spectrum of a periodic Hamiltonian to trace the trajectory of the underlying Floquet eigenstates, thus extending coherence times ($T_2$) and inheriting the innate robustness of the DD protocol.
\par
In this work, we experimentally demonstrate this adiabatic DD protocol, AdPulse, to achieve coherent polarization transfer from NV electronic spins to nearby, coupled $^{13}\text{C}$ nuclear spins.
We detect the coherent polarization dynamics directly using electron spin resonance-like free induction decay spectra at room temperature without the need for RF pulses or population transfer from the nuclear spin to the electronic spin. We also compare the performance of AdPulse directly to its diabatic counterpart, highlighting regimes where adiabatic methods are more robust and effective. Additionally, we numerically investigate the polarization of $^{13}\text{C}$ clusters, providing further insight into the potential applications of adiabatic pulsed operation in quantum systems.

\paragraph{Hybrid central spin system and Floquet treatment:} \label{theory}
\begin{figure}[h!]
	\includegraphics[trim= 0cm 0.1cm 0cm 0cm, clip=true,width=1\columnwidth]{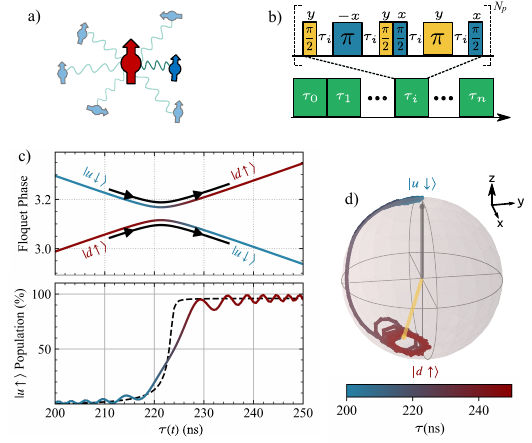}
	\caption{Illustration of the NV and AdPulse sequence. (a) Schematic of the hybrid electron (red) - nuclear (blue) spin system (b) Implementation of the AdPulse sequence. (c) Top: Floquet spectrum of the target state's trajectory under AdPulse protocol. Bottom: Adiabatic state transfer dynamics via the avoided crossing, where the dashed line (analytical LZ formula, Eq.\eqref{Eq: LZ}) captures the adiabatic limit. (d) Bloch sphere representation showing the oscillatory orbits around the LZ prediction axis.
		\label{fig1}}
\end{figure}
We consider a central electronic spin in a lattice (Fig.\ref{fig1}(a)) under an external magnetic field, $B_0$, which is aligned along the $z$-axis. The central spin is coupled to a local nuclear spin cluster of size $N_\mathrm{nuc}$ which can be approximated as a purely dephasing hyperfine interaction, assuming large energy level separation of central spin states. Working in the rotating frame of MW driving resonant with the electronic spin transition, the system Hamiltonian can be written as:
\begin{equation}
\hat{H}(t) = \sum_n^{N_\mathrm{nuc}} \gamma_n B_0 \hat{I}_z^{(n)}  + \hat{S}_z\sum_n^{N_\mathrm{nuc}} \mathbf{A}^{(n)}\cdot \hat{\mathbf{I}}^{(n)} + \hat{H}_c(t),
\label{eq1}
\end{equation} 
where $\mathbf{A}^{(n)} = A^{(n)}_x\mathbf{e}_x  + A^{(n)}_z\mathbf{e}_z$ ($\gamma_e B_0 >> A^{(n)}_x, A^{(n)}_z$) is the hyperfine coupling vector and $\gamma_{e},\gamma_{n}$ are the gyromagnetic ratio of the electronic and $n^\mathrm{th}$ nuclear-spin, respectively, with $\hat{S}_j$ and $\hat{I}^{(n)}_j$ as their corresponding spin operators. We adopt notation $\mathbf{e}_i$ for cartesian unit vectors. Coherent control of the electronic spin is achieved through the control Hamiltonian $\hat{H}_c(t) = \Omega_x(t)\hat{S}_x + \Omega_y(t)\hat{S}_y$, where $\Omega_j(t)$ are the time-dependent waveforms along the \(j\)-axis. For many pulsed control schemes, MW $\pi$-pulses are applied to periodically invert the electronic spin.
When the inverse of the pulse spacing, $\tau$, is on resonance with the precession frequency of the nuclear spins coupled to electronic spin, their dynamics become correlated. For common DD protocols like XY-8 this occurs when the pulse spacing 
$\tau_r^{(n)} = K\pi/\omega^{(n)}_I$ for odd integer $K$. Here, the net nuclear precession frequency is given by $\omega^{(n)}_I = \sqrt{(\omega_L - A^{(n)}_z/2)^2 + (A^{(n)}_x/2)^2}$  for the $n^{th}$ nuclear spin with Larmor frequency $\omega_L = \gamma_n B_0$.

Control protocols have been developed for selective spin state control allowing for DNP \cite{abragam1978principles}. Using these protocols, thermally polarised (low purity mixed states) nuclear spins can be initialised into higher statistically polarised states. 
A recently developed pulsed DNP protocol, PulsePol, creates a flip-flop interaction between the electronic and nuclear spin using a series of orthogonal phase pulses, and has been used to polarize nuclear spins in diamond \cite{schwartz2018robust}. The electronic spin is resonant with the $n^{th}$ nuclear spin at the resonant pulse spacing $\tau_r^{(n)} \simeq K\pi/4\omega_I^{(n)}$, for odd $K$. For $K = 3$, the flip-flop rate is maximal and the effective Hamiltonian is 
\begin{equation}
\hat{H}_\mathrm{eff} = g^{(n)}_3(\hat{S}_+\hat{I}^{(n)}_- + \hat{S}_-\hat{I}^{(n)}_+)/2,
\label{eqflipflop}
\end{equation}
where $g_3 = A^{(n)}_x(2 + \sqrt{2})/6\pi$
\cite{schwartz2018robust}.

The majority of pulsed protocols are temporally periodic, allowing us to employ \textit{Floquet theory} to analyze the resulting dynamics. \textit{Floquet's theorem} states that for a temporally periodic Hamiltonian $\hat{H}(t + T) = \hat{H}(t)$, the eigenstates of the system can be written $|\psi_l(t)\rangle = \exp[-i\epsilon_l t]|\chi_l(t)\rangle$ with quasi-energies $\epsilon_l$ and Floquet states $|\chi_l(t)\rangle$, which are time-periodic with $|\chi_l(t + T)\rangle = |\chi_l(t)\rangle$. At stroboscopic times $N_pT$, these Floquet states are eigenstates of the system. Under the single-period evolution operator $\hat{U}(T)$ they therefore only gain a phase such that $\hat{U}(T)|\chi_l\rangle = e^{-i\varepsilon_l(T)}|\chi_l\rangle$ with $|\chi_l\rangle = |\chi_l(0)\rangle$, where $\varepsilon_l(T)$ are known as Floquet phases. These Floquet phases can be solved analytically using $\varepsilon_l = \tan^{-1}(\mathrm{Im}\lambda_l/\mathrm{Re}\lambda_l)$, however they may also be found by numerically by diagonalizing the single period evolution operator. Analysis of dynamics is then performed using \textit{Floquet spectroscopy} \cite{lang2015dynamical}, in which the calculated Floquet phases are plotted for different periodicities $T$ (Fig.\ref{fig1}(c)) - analogous to spectral eigenenergy analysis, where resonances between coupled states appear as avoided crossings in the spectrum. As with eigenenergy spectra, trajectories of the Floquet states in the Floquet spectrum of a particular DD protocol can be adiabatically traced using the AdPulse family of protocols. 

As shown in Fig.\ref{fig1}(b) and in reference  \cite{whaites2022adiabatic}, the procedure for AdPulse is as follows: we start by applying our chosen DD sequence $N_p$ times with an initial pulse spacing $\tau_{0}^{(n)} \ll \tau_r^{(n)}$ (for a specific $K$); next, we apply the same DD sequence with a pulse spacing increased by the step size $\delta\tau \ll \tau_r^{(n)}$ leading to the new pulse spacing $\tau_{0}^{(n)} + \delta\tau$; we incrementally repeat this procedure $m$ times until a step size $\tau_f^{(n)} \gg \tau_r^{(n)}$ is reached and the sweep has passed through the protocol resonance, or avoided crossing in the Floquet spectrum. The range of the sweep is then $\Delta\tau = \tau_f^{(n)} - \tau_0^{(n)}$. As with conventional adiabatic state transfer, if the sweep is linear, then \textit{Landau-Zener} (LZ) theory can be used to construct a semi-classical closed form expression for the sweep dynamics \cite{zener1932non,vitanov1999transition}. The population remaining in the initial state after a sweep through a two-state avoided crossing is given by
\begin{equation}
    P_0(\tau) = e^{-2\pi\Gamma_0F(\tau;\tau_0)}
    \label{Eq: LZ}
\end{equation} where $\Gamma_0$ is the Landau-Zener exponent and $F(\tau;\tau_0) = \frac{1}{\pi}[\arctan(\Phi(\tau)) - \arctan(\Phi(\tau_0))]$ defining $\Phi(\tau) = 2\Delta(\tau)\sqrt{\dot{\Delta}/4\pi}$.\cite{whaites2022adiabatic} For the flip-flop Hamiltonian of Eq.\ref{eqflipflop}, $\Gamma_0 = \frac{1}{2\pi}\frac{\tau_r^3A_x^2}{\beta^2\delta\tau}$, $\Delta(\tau) = \omega_I - \omega$ defining $\beta = 3\pi/(2 + \sqrt{2})$ and $\dot{\Delta} \simeq \frac{3\pi}{16\tau_r^3}\frac{\delta\tau}{N_p}$. The protocol frequency is defined $\omega = 2\pi K/T$
Fig.\ref{fig1}(c) shows the semi-classical analytical form of an AdPulse sweep through a two-level avoided crossing, corresponding to Eq.\eqref{Eq: LZ}(dashed curve), alongside full numerical solution (solid curves). We visualize the full dynamics of the AdPulse sweep on the Bloch sphere Fig.\ref{fig1}(d), showing a precession around the predicted LZ limit. 
\begin{figure}[h!]
	\includegraphics[trim= 0.5cm 9cm 0.5cm 4cm, clip=true,width=1\columnwidth]{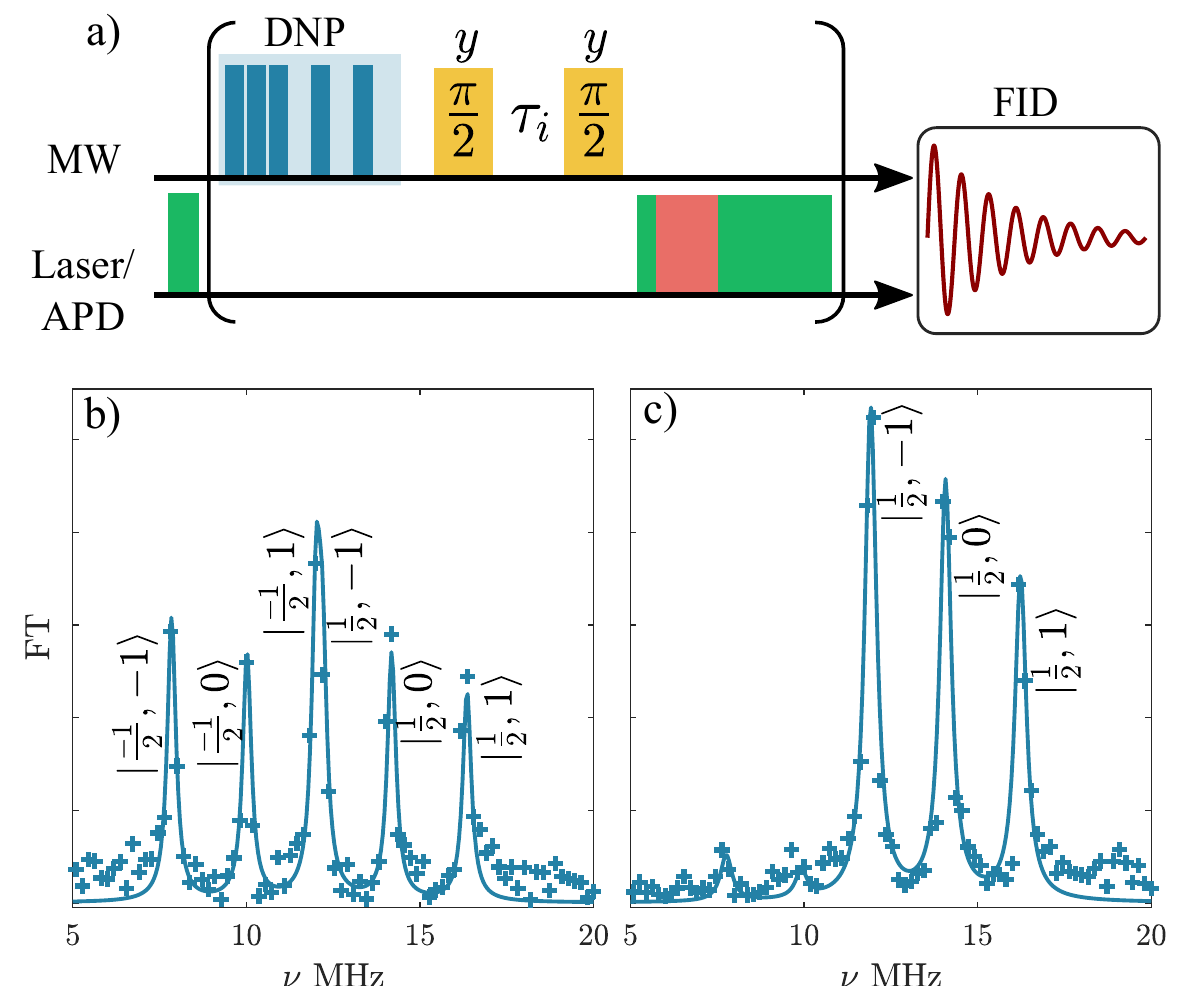}
	\caption{Nuclear spin polarization, detection, and re-initialization. (a) Schematic for the pulse sequence used. Following the DNP protocol (in blue) the detection sequence comprises of a pulsed Ramsey scheme (in yellow). The laser (green) and avalanche photo-diode (APD, red) provide electron spin initialization and readout. (b)-(c) Fourier transform (FT) of the experimental FID signals (markers) for a NV-A.
 		\label{fig2}}
\end{figure}


\paragraph{Experimental methods:}
We use NV centers in diamond coupled with $^{13}$C nuclear spins as our central spin system. The NV center triplet ground state ($S=1$) has a 2.87\,GHz zero-field splitting between the $m_s=0$ and $m_s=\pm1$ states. We apply an external magnetic field $B_0$ along the NV axis, which lifts the $m_s=\pm1$ degeneracy \cite{gruber1997scanning,jelezko2002single}. Selective microwave (MW) driving of the $|0\rangle\rightarrow|m=-1\rangle$ transition isolates an effective two-level system, described by the rotating frame Hamiltonian in Eq.\ref{eq1} \cite{App-1}. We use Gaussian MW pulses of 16\,ns duration, with the $\pi/2$ pulse amplitude set to half that of the $\pi$ pulses. After laser initialization of the NV, a DNP protocol is applied via MW control, and $^{13}$C polarization is quantified by a Ramsey sequence \cite{App-1},  followed by a long laser pulse that both detects and resets the electronic spin state, as shown in Fig.\ref{fig2}(a). This long laser pulse also depolarizes the $^{13}$C nuclei, due to strong excited-state hyperfine coupling which leads to polarization leakage through the excited state level anti-crossing (ESLAC) \cite{gaebel2006room, fuchs2008excited, jacques2009dynamic, gali2009hyperfine}. We utilize this laser-induced depolarization to generate a thermal $^{13}$C nuclear spin state at the start of each measurement \cite{App-1}.

Repeating the detection sequence for varying Ramsey free evolution intervals yields a time-domain free induction decay (FID). Fourier transforms of the FID signals for NV-A (see \cite{App-1}) coupled to a nearby $^{13}$C and the host $^{14}$N, shown in Fig.\ref{fig2}(b,c), reveal an ESR-like spectrum with six hyperfine peaks of the NV $|-1\rangle$ state, labeled as $|m_C, m_N\rangle$. Overlap between the $|{-1/2,1}\rangle$ and $|{1/2,-1}\rangle$ levels reduces the observed peaks to five. We determine $^{13}$C polarization by comparing the normalized areas of the peaks corresponding to the $m_C=-1/2$ (i.e., $|{-1/2,-1}\rangle+|{-1/2,0}\rangle+|{-1/2,1}\rangle$) and $m_C=1/2$ (i.e., $|{1/2,-1}\rangle+|{1/2,0}\rangle+|{1/2,1}\rangle$) manifolds, effectively tracing out the $^{14}$N state (see \cite{App-1} for details).
Fig.\ref{fig2}(b) shows a maximally mixed $^{13}$C spin, while Fig.\ref{fig2}(c) illustrates an 91\% polarization achieved using an AdPulse sweep. This Ramsey based detection with optical re-initialization thus directly quantifies nuclear spin polarization.
 
\paragraph{Adiabatic pulsed polarization of single spins:}
\begin{figure}[h!]
	\includegraphics[trim= 5cm 3.6cm 4.5cm 2cm, clip=true,width=1\columnwidth]{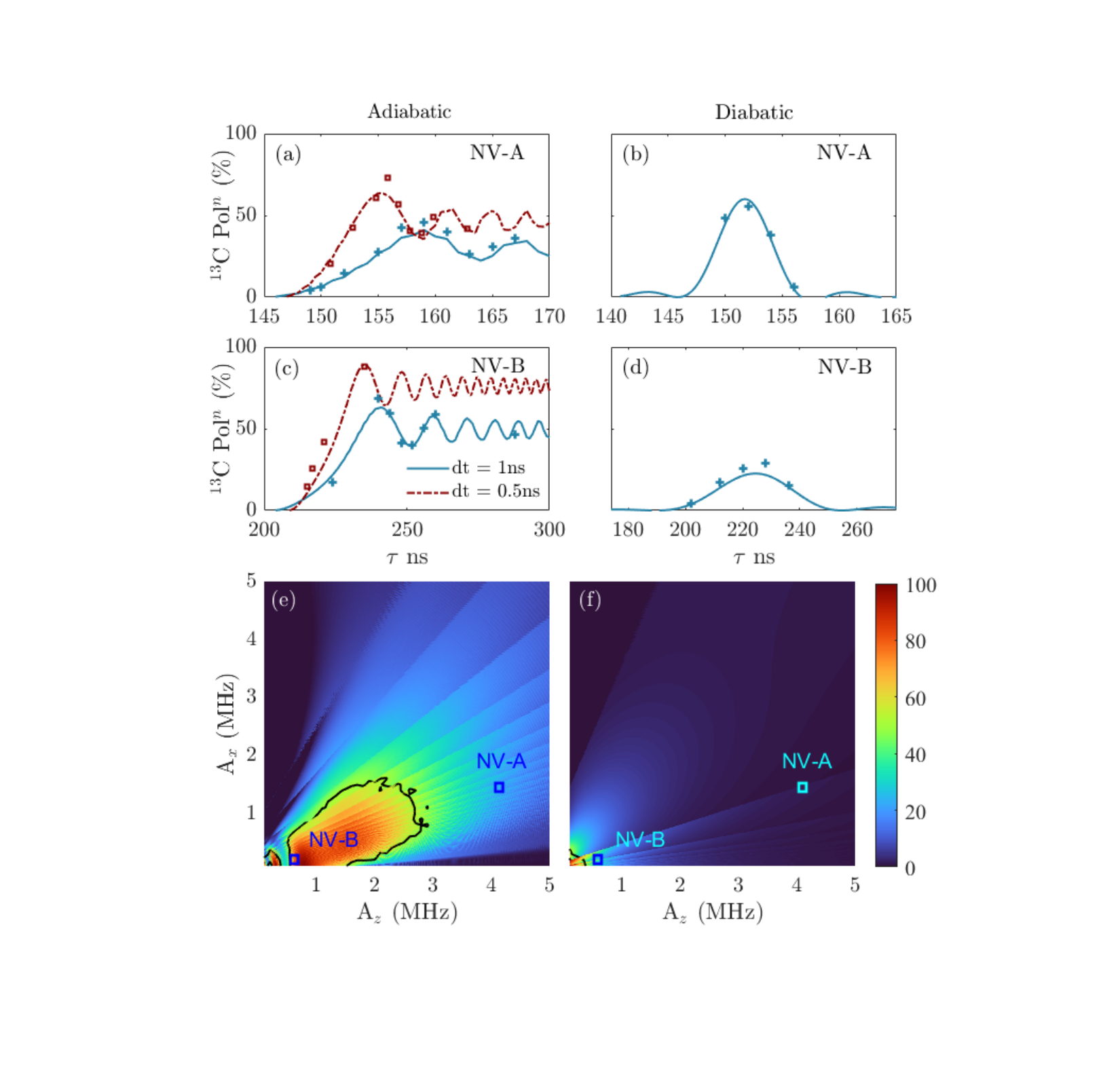}
	\caption{Experimental and simulated data showing advantage of using adiabatic over diabatic operation. Panels (a) and (b) compare simulation (curves) and experiment (markers) for $^{13}$C polarization transfer on NV-A ($A_z=4.12$ MHz, $A_x=1.42$ MHz) under AdPulse sweeps with $\delta\tau=1\,\mathrm{ns}$ and $0.5\,\mathrm{ns}$, and under PulsePol with $N_p=16$, respectively for $K = 3$. Panels (c) and (d) show analogous results for NV-B ($A_z=607$ kHz, $A_x=137$ kHz), for $K = 1$. Panels (e) and (f) display surface plots of $^{13}$C polarization (color map in $\%$) for different hyperfine coupling components for AdPulse and PulsePol (for $K=3$), where PulsePol uses the optimal scheme presented in \cite{schwartz2018robust}. Black contour lines mark regions with over 50\% polarization transfer.
		\label{fig3}}
\end{figure} 
We focus here on the adiabatic realization of the PulsePol sequence, which enables the initialization of nuclear spins using a central electronic spin. The PulsePol DNP protocol has been extensively used with NV centers to polarize individual spins, spin clusters, and nuclear spin baths within the diamond lattice \cite{randall2021many,blinder202413}, including work targeting spins outside the diamond lattice \cite{healey2021polarization}. We compare this diabatic DNP protocol with its adiabatic counterpart, AdPulse, focusing on polarization transfer efficiency and robustness using two NV centers (NV-A and NV-B) in natural abundance $^{13}$C Type IIa diamond, each coupled to a single distinct $^{13}$C with their hyperfine components representing extremes in hyperfine coupling strength \cite{App-1}. Fig.\ref{fig3}(a) and (b) show polarization dynamics of a $^{13}$C nuclear spin coupled to NV-A,  with AdPulse and PulsePol DNP protocol, respectively, while Fig.\ref{fig3}(c) and (d) display results for a $^{13}$C coupled to NV-B. AdPulse is implemented with an initial pulse spacing of $\tau_0^{(1)} = \tau_r^{(1)} - 5$ ns
and step sizes $\delta \tau = 1$ ns (blue markers) and $0.5$ns (red markers), whereas PulsePol uses $N_p=16$ with $K = 3$ for NV-A and $K = 1$ for NV-B. 
For NV-B we choose the first resonance, $K = 1$, to ensure comparable operational times between the two NVs, owing to the difference in their hyperfine couplings. 
Numerical simulations (curves) reproduce experimental features, including oscillations following the level anti-crossing for the case of AdPulse. Our direct detection protocol tracks $^{13}$C polarization coherently over time at room temperature without recourse to RF-based nuclear manipulations. AdPulse demonstrates that greater adiabaticity—achieved with smaller step sizes ($\delta\tau$)—yields higher polarization transfer level and efficiency. For AdPulse an optimal $\tau_f^{(n)}$ exists for maximum polarization transfer, which can be determined numerically or by examining Floquet eigenstate dynamics. Remarkably, sweeping past this optimum time results in an oscillatory polarization transfer to a finite saturation polarization or LZ limit (shown in \ref{fig1}(c)), which improves with increased adiabaticity. This saturation persists until the next resonance, typically lasting up to hundreds of nanoseconds.
PulsePol, in contrast, exhibits a narrower polarization transfer window (see \ref{fig3}(b) and (d)). The resonance width for AdPulse is approximately 10–20 times broader than the PulsePol, enhancing its resilience to uncertainties in $^{13}$C coupling parameters. Furthermore, AdPulse inherits robustness to detuning and Rabi errors from PulsePol, as discussed in \cite{schwartz2018robust}.

Next we assess the performance of AdPulse and PulsePol protocols across a wide range of hyperfine couplings. For this, we adopt the optimal PulsePol parameters at the second resonance ($K = 3$), proposed in \cite{schwartz2018robust}, individually tuned for each coupling parameter considered. Similarly, for AdPulse we use the analytical expressions of $\delta \tau$ and total sweep time, given in \cite{App-1}, symmetric about $\tau_r^{(n)}$.
Simulated results, shown in Fig.\ref{fig3}(e)-(f), take into account the finite Gaussian pulses of \(16 \, \mathrm{ns}\) duration. 
Polarization using PulsePol shows good performance in the low coupling regime but the performance dramatically drops at larger values of parallel and perpendicular component of the hyperfine coupling.
In contrast, AdPulse efficiently targets a wide range of hyperfine couplings. This can be seen clearly in the 50$\%$ contour lines shown in \ref{fig3}(e)-(f). While PulsePol parameters can be fine-tuned to maximize polarization for specific coupling strengths, magnetic fields, or pulse durations, its efficiency diminishes under parameter uncertainties. AdPulse, however, maintains robust and efficient performance in such scenarios, providing a significant advantage in practical applications. 

\par
\begin{figure}[h!]
	\includegraphics[trim= 0cm 0cm 0cm 0cm, clip=true,width=1\columnwidth]{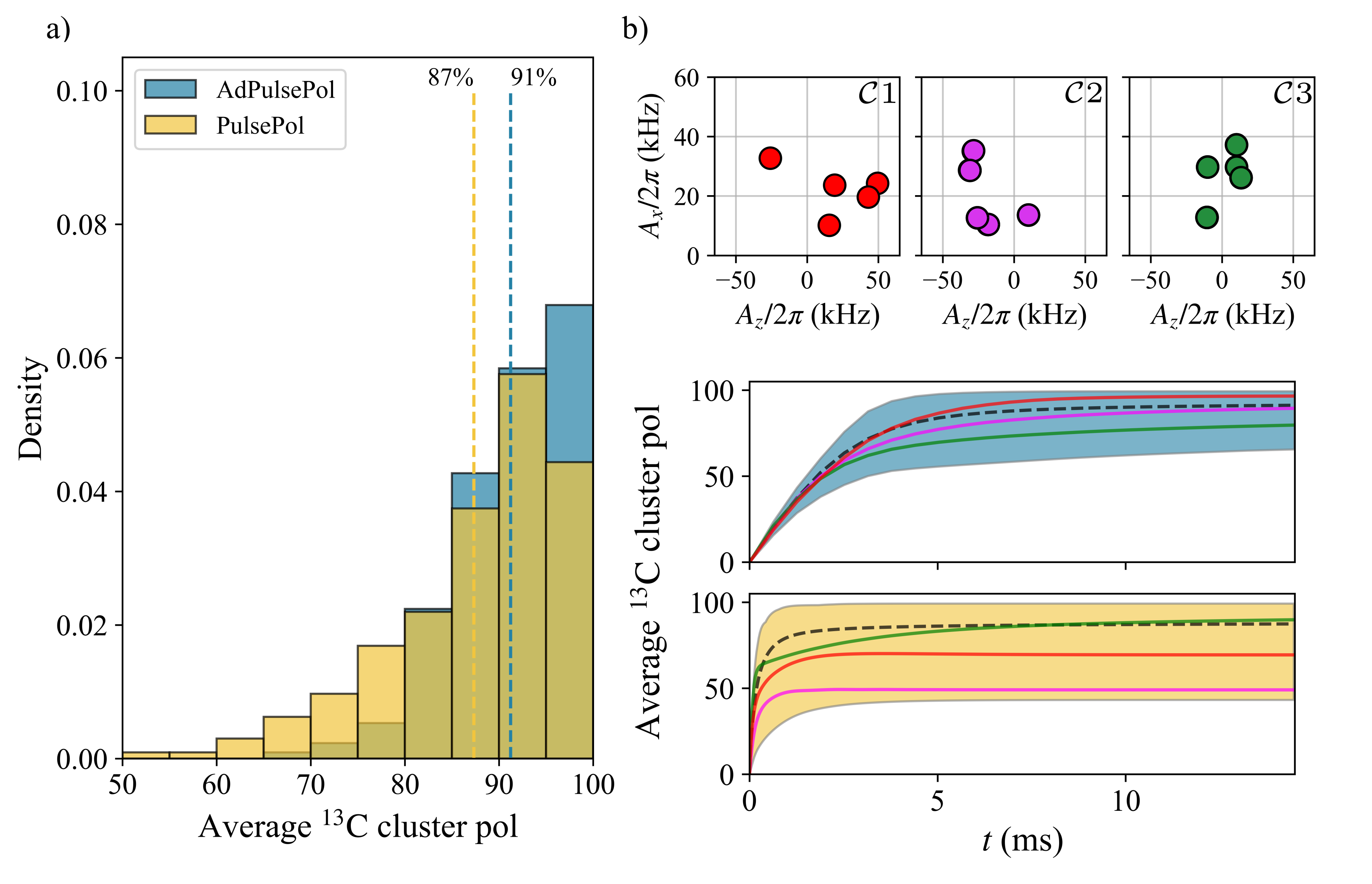}
	\caption{Comparison of AdPulse and PulsePol protocols for cluster hyperpolarization. Polarization of a 5-spin cluster is simulated using AdPulse with $N_p=2$, $K=3$, $\tau_r \simeq 1497\,\mathrm{ns}$, $\Delta\tau = 250\,\mathrm{ns}$, and sub-adiabatic $\delta\tau = 5\,\mathrm{ns}$, and using PulsePol with $N_p=4$ and $\tau = \tau_r$. In both cases the NV center is repeatedly reinitialized ($R=25$ for AdPulse and $R=476$ for PulsePol), yielding an equal operating time of $t = 15\,\mathrm{ms}$ (assuming $\sim6\,\mu\mathrm{s}$ per re-initialization). Panel (a) presents a histogram of the average polarization of 850 randomly generated 5-spin clusters ($^{13}\mathrm{C}$ couplings $A_{x,z}/2\pi \in [10,60]\,\mathrm{kHz}$). Panel (b) highlights three specific clusters ($\mathcal{C}1$, $\mathcal{C}2$, $\mathcal{C}3$); the lower subpanel shows the polarization buildup for each protocol, with the ensemble average indicated by a dashed line.
		\label{fig4}}
\end{figure}

  
\paragraph{Adiabatic pulsed polarization of spin clusters:}
An important task for nuclear spin polarization protocols is the initialisation of multi-qubit clusters of 5–10 $^{13}\mathrm{C}$ spins for applications in nuclear spin memories and many-body quantum simulation. These clusters typically exhibit weak hyperfine couplings ($A_{x,z} \in [10,60]$ kHz), and require \textit{hyperpolarization} for simultaneous polarization. Fig.\ref{fig4} shows a numerical simulation comparing this process for AdPulse and PulsePol, over a total polarisation time of $15\,\mathrm{ms}$.
For AdPulse, a broad sweep of $\Delta\tau = 250\,\mathrm{ns}$ is executed through the Larmor resonance ($\tau_r = 3\pi/4\omega_L \simeq 1500\,\mathrm{ns}$) with a sub-adiabatic step size of $\delta\tau = 5\,\mathrm{ns}$. This sweep range is chosen to suit a broad range of hyperfine couplings. Each AdPulse sweep cycle lasts $t_\mathrm{cyc} \simeq 700\,\mu\mathrm{s}$, which is within the $T_2$ limit for single NV experiments with over 1000 MW pulses \cite{abobeih2018one}. After each sweep, the NV is reinitialized; only $R = 25$ reinitializations are needed for the polarization to saturate. In contrast, the protocol for diabatic polarization with PulsePol uses small packets ($N_p = 4$) at $\tau = \tau_r$, leading to approximately $R = 476$ re-initializations within the total operation time $t_\mathrm{ex} = 15\,\mathrm{ms}$. This approach appears to be optimal for hyperpolarization, where for large $N_p$ polarization tends to randomize per cycle for large clusters (see \cite{App-1}).
To capture many-body effects, we simulate the average polarization for 850 unique 5-spin clusters with hyperfine couplings $A_{x,z}/2\pi \in [10,60]$ kHz. To limit finite pulse effects, we initialize the $^{14}\mathrm{N}$ host into the $m = 0$ state, which can be done experimentally. The histogram in Fig.\ref{fig4}(a) shows that after $t_\mathrm{ex} \simeq 15\,\mathrm{ms} \ll T_1$ of the $^{13}\mathrm{C}$, AdPulse yields on average about 4\% greater cluster polarization while requiring roughly 20 times fewer NV re-initializations than PulsePol. This reduced re-initialization may additionally be beneficial in mitigating laser-induced depolarization of $^{13}\mathrm{C}$ spins, suggesting that the use of AdPulse could further improve cluster polarization.

For three randomly generated clusters, labeled $\mathcal{C}1$, $\mathcal{C}2$, and $\mathcal{C}3$, Fig.\ref{fig4}(b) displays their hyperpolarization dynamics along with the corresponding hyperfine couplings. While PulsePol initially achieves moderate polarization levels (approximately 60\% for clusters $\mathcal{C}1$ and $\mathcal{C}2$), their dynamics soon saturate. In contrast, AdPulse steadily increases the polarization, ultimately reaching over 90\% for the two clusters. This improved performance is likely due to the variance in the nuclear resonances and the spread of $A_z$ values, which render a fixed pulse spacing $\tau$ suboptimal for off-resonant polarization \cite{whaites2024hyperpolarisation}. 
AdPulse overcomes this challenge by adiabatically sweeping through all resonances, thereby addressing each spin individually. Moreover, nuclear spin pairs $i$ and $j$ satisfying $|A_z^i - A_z^j| \ll A_x^i, A_x^j$ can form spin-pair dark states that have been shown to reduce or even block polarization efficiency \cite{villazon2021shortcuts,randall2021many}. In Fig.\ref{fig4}(b), such dark states are evident in clusters $\mathcal{C}1$ and $\mathcal{C}2$, where AdPulse prevents early saturation by mitigating dark state formation. In contrast, for cluster $\mathcal{C}3$, with $A_z$ values closely grouped around zero, PulsePol achieves a higher average polarization (approximately 90\%) compared to AdPulse (approximately 75\%).


\paragraph{Discussion and conclusions:}
Our results establish that integrating adiabatic sweeps within a dynamical decoupling (DD) framework (AdPulse) provides robust and efficient nuclear spin polarization. The nuclear spin ensemble adiabatically follows the evolving Floquet eigenstates of the underlying DD Hamiltonian. This adiabatic passage results in coherent polarization transfer, as demonstrated by the FID-based direct readout, which shows excellent agreement with numerical simulations and theoretical predictions based on Landau-Zener (LZ) analysis. Although the LZ model captures the overall adiabatic limit, full quantum simulations reveal additional oscillatory dynamics—visualized as Bloch sphere precessions.
A head-to-head comparison with the established PulsePol protocol reveals several key advantages of its adiabatic counterpart, AdPulse. In single-$^{13}\mathrm{C}$ experiments, AdPulse achieves higher polarization efficiency with increasing adiabaticity, 
yielding extended resonance windows approximately 10–20 times wider than those of PulsePol. This enhances robustness against uncertainties in the hyperfine coupling parameters, pulse timings (from fluctuations in the ambient physical parameters), and pulse imperfections.
Furthermore, the hyperpolarization numerical simulations on $^{13}\mathrm{C}$ clusters demonstrate that AdPulse yields on average a $\sim$4\% improvement in cluster polarization—all while requiring significantly fewer NV reinitialization cycles compared to the diabatic PulsePol approach. Although specific clusters with tightly grouped $A_z$ values may occasionally favor PulsePol, on average the broader operational window of AdPulse offers clear performance benefits, particularly relevant for applications involving multiple nuclear spin memory qubits or spin-based many-body quantum simulation. On the other hand, in cases where the coherence time $T_2$ of the NV center is limited, rapid polarization using PulsePol may be preferred. Alternatively, AdPulse sweep times may be reduced by optimizing sweep velocities or implementing shortcuts to adiabaticity \cite{guery2019shortcuts}. In fact on a single cluster level, sweep times may be significantly reduced by choosing a smaller range $\Delta\tau$ which is more suited to its specific resonance distribution.
Overall, the enhanced robustness and extended resonance behavior of AdPulse underscore its potential for scalable qubit initialization and quantum sensing, where environmental uncertainties and pulse errors are inherent.
This adiabatic approach could also be applied to other pulsed sequences for robust operations on single or multiple nearby spins in hybrid electron-nuclear systems. 

\paragraph{Acknowledgment:}
We gratefully acknowledge the insightful discussions on experimental design and hardware implementation, as well as the valuable suggestions for refining the manuscript, provided by Toby Mitchel and Jack Hart. We also thank Mete Atature and Dirk Englund for support with sample preparation and Daniel Louzon for help with the experimental hardware. This work was supported by the EPSRC Grant EP/V049704/1 and by the Royal Society through a University Research Fellowship held by H.S.K. O.T.W acknowledges support from DTP studentship grant EP/R513143/1 of the Engineering and Physical Sciences Research Council (EPSRC) as well as Quench project that has received funding from the European Union's Horizon Europe -- The EU Research and Innovation Programme under grant agreement No 101135742.

\renewcommand{\theequation}{A\arabic{equation}}
\setcounter{equation}{0}  

\bibliographystyle{ieeetr}
\bibliography{lib}

\end{document}